\documentclass[epj]{svjour}
\usepackage{graphics}
\begin{document}
\title{Heavy Quarkonia from Classical SU(3) Yang-Mills Configurations}
\author{R. A. Coimbra \and O. Oliveira 
}                     
%
%
\institute{Centro de F\'{\i}sica Computacional, Universidade de Coimbra,
           3004 516 Coimbra, Portugal}
\date{Received: date / Revised version: date}
%
\abstract{A generalized Cho-Faddeev-Niemi ansatz for SU(3) Yang-Mills is
investigated. The corresponding classical field equations are solved for its
simplest parametrization. From these solutions it is possible to define a
confining central non-relativistic potential used to study heavy quarkonia. The
associated spectra reproduces the experimental spectra with an error of less
than 3\% for charmonium and 1\% for bottomonium. Moreover, the recently
discovered new charmonium states can be accomodate in the spectra, keeping the
same level of precision. The leptonic width show good agreement with the
recent measurements. The charmonium and bottomonium E1 electromagnetic 
transitions widths are computed and compared with the experimental values.
\PACS{
      {12.39.Pn}{Potential models}   \and
      {12.38.Lg}{Other nonperturbative calculations} \and
      {12.38.-t}{Quantum chromodynamics}
     } 
} 
\maketitle
%
\section{Introduction}
\label{intro}


Charmonium and bottomonium are essentially non relativistic systems. Their
dynamics can be understood in terms of a potential. Ideally, we would be able 
to derive such a potential directly from QCD. However, the confining 
potentials currently used to describe such systems, although being related to 
the fundamental theory, are not directly derived from QCD.

In this paper, we report on a class of classical solutions of the Yang-Mills
theory, computed after the introduction of a generalized Cho-Faddeev-Niemi 
\cite{FaNi99,Cho}
ansatz for the gluon field. The solutions allow for a definition of a
confining potential which is then used to study charmonium and bottomonium 
spectra, their leptonic decays and charmonium electromagnetic transitions. For
interquark distance between 0.2 fm to 1 fm, the new potential is essentially 
the lattice singlet potential. For larger distances it grows exponentially 
with the interquark distance and for smaller distances is coulomb like. In 
what concerns the spectra, the new potential is able to reproduce
charmonium and bottomonium spectra with an error of less than 3\% and 1\%,
respectively, including the recently discovered new charmonium states
$X(3872)$, $X(3940)$, $Y(3940)$, $Z(3940)$ and $Y(4260)$. If the spectra
is reasonable well described, the hyperfine splittings $^3S_1 - ^1S_0$ turn 
out to be to small. This can be due to a missing scalar confining potential or
the need for including contributions from other configurations, namely those
related with the short distance behaviour of the theory. In what concerns the
leptonic decays and the charmonium electromagnetic transitions, the new
potential is able to describe well the experimental data. Part of this work
is described in \cite{OlRi06}.

\section{Classical Configuration and a Heavy Quark Potential}
\label{sec:1}

In the following we use the notation of \cite{Hu92}. Following a procedure
suggest in \cite{Cho80}, we introduce a real covariant constant field $n^a$,
\begin{equation}
 D_\mu n ^a   ~ =  ~ \partial_\mu n^a \, + \,  g f_{abc}  n^b A^c_\mu ~ = ~ 0
  \, . 
 \label{na}
\end{equation}
Multiplying this equation by $n^a$ it cames out that $n^a n^a$ can be chosen to
be one. Defining the color projected field $C_\mu$ such that
\begin{equation}
 A^a_\mu ~ = ~ n^a \, C_\mu ~ + ~ X^a_\mu \, ,
 \hspace{0.3cm} n^a X^a_\mu = 0 \, ,
\end{equation}
and replacing this definition in (\ref{na}), $X^a_\mu$ can be related to $n^a$
- see \cite{OlRi06} for details. For the simplest parametrization of 
$n^a = \delta^{a1} \sin \theta + \delta^{a2} \cos \theta$ and for the simplest
gluon configuration, the classical field equations in Landau gauge becomes
\begin{equation}
 \partial^\mu \partial_\mu \theta ~ = ~ 0 \, ,
 \label{euler}
\end{equation}
with $g A^a_\mu = - \delta^{a3} \partial_\mu \theta$.
Note that there are no boundary conditions for $\theta$. Equation (\ref{euler})
can be solved by the usual method of separation of variables. The solution
considered here being
\begin{eqnarray}
 A^3_0      &  =  &
   \Lambda \Big( e^{\Lambda t} - b_T e^{- \Lambda t} \Big) V_0 (r),  \\
 \vec{A}^3  & = & 
    - \Big( e^{\Lambda t} + b_T e^{- \Lambda t} \Big) 
     \nabla V_0 (r), \\
  V_0 (r)   & = &
    A \frac{\sinh ( \Lambda r)}{r} + 
    B \frac{e^{- \Lambda R}}{r} \, .
\end{eqnarray}
Assuming that the spatial part of $A^3_0$ can be identified with a non 
relativistic potential for heavy quarkonia (see \cite{OlRi06} for details), the
squared difference between $V_0$ and the lattice singlet potential integrated
between 0.2-1 fm was minimized to defined the various parameters; 
for $r$ in MeV$^{-1}$, 
A = 11.254, B = -0.701, $\Lambda$ = 228.026 MeV. The difference
between $V_0$ and the singlet potential is less than 50 MeV for 
$r \in [0.2 - 1]$ fm. In the following, the spin-dependent potential 
assumes that the interaction is pure vectorial type and that $V_0$ describes 
the interaction between the heavy quarks.

\subsection{Charmonium and Bottomonium Spectra}
\label{sec:2}

The charm quark mass, $m_c = 1870$ MeV, was adjust to reproduce the 
experimental value of the $\chi_{c2} (1P) - J / \psi (1S)$ mass difference. 
The bottom quark mass, $m_b = 4185.25$ MeV, was defined to reproduce the 
experimental value for the $\Upsilon (2S) - \Upsilon (1S)$ mass difference.

%
\begin{table}
\caption{Charmonium Spectra. The right columns report the difference, in MeV,
between theory and experimental numbers.}
\label{charm}       
\begin{tabular}{llr|llr}
\hline\noalign{\smallskip}
 Theo  & Exp   &      & Theo  & Exp   &  \\
 (MeV) & (MeV) &      & (MeV) & (MeV) &  \\
\hline\noalign{\smallskip}
 \multicolumn{3}{l|}{$J^{PC} = 0^{-+}$}  &    \multicolumn{3}{l}{[$1P$]}    \\
 3075 & $\eta_c (1S)$        & 95       &    3556 & $\chi_{c2} (1P)$  & 0  \\
 3632 & $\eta^\prime_c (2S)$ & -5       &    3462 & $\chi_{c1} (1P)$  & -49 \\
 4135 & $Y(4260)$            & -125     &    3372 & $\chi_{c0} (1P)$  & -43 \\
 4638 & ---  & ---                      &    3478 & $h_c (1P)$        & -48 \\
\hline\noalign{\smallskip}
 \multicolumn{3}{l|}{$J^{PC} = 1^{--} ~$ [$^3S_1$]} &  
 \multicolumn{3}{l}{$J^{PC} = 1^{--} ~$ [$^3D_1$]} \\
 3097 & $J / \psi (1S)$  & 0    &      &                &    \\
 3659 & $\psi (2S)$      &  -27 & 3688 & $\psi (3770)$  & -83 \\
 4164 & ---              & ---  & 4155 & ---            & --- \\
\hline\noalign{\smallskip}
\multicolumn{6}{l}{$M [ J / \psi (1S) - \eta_c (1S)] = 
                    \left. 22 \right|_{\mbox{Theo}} =
                    \left. 117 \right|_{\mbox{Exp}}$ MeV} \\
\multicolumn{6}{l}{$M [ \psi (2S) - \eta^\prime_c (2S)] 
          \hspace{0.38cm} = 
                    \left. 26 \right|_{\mbox{Theo}} = ~
                    \left. 48 \right|_{\mbox{Exp}}$ MeV} \\
\hline\noalign{\smallskip}
\end{tabular}
\end{table}

In table \ref{charm} we report the low lying charmonium states. Note that the 
masses were shifted to reproduce the $J/ \psi (1S)$ experimental value. 
The third columns of table gives, in MeV, the differences between the 
theoretical prediction and the experimental values taken from \cite{PDG}. 
In what concerns the spectra, there is good agreement between the theoretical
spectra and the measured values. The only particles which don't fit well in
spectra are $\psi (4040)$ and $\psi (4415)$. In what concerns these two states,
the experimental information is quite scarse and the particle data book 
comments that the ``interpretation of these states as a single resonance is 
unclear because of the expectation of substantial threshold effects in this 
energy region''. Curiously, both particle masses are essentially the sum of 
$J / \psi$ with $J^{PC} = 0^{++}$ known mesons.

In what concerns the new charmonium states, the theoretical predicitions which
are close to the experimental values are: 
$Z(3930)$, mass of $3929 \pm 5 \pm 2$ MeV, $J = 2$, is a $1^3F_2$ state with 
3932 MeV or a $2^3P_2$ state with 4048 MeV ; $X(3940)$, mass of 
$3942 \pm 11 \pm 13$ MeV, and $Y(3940)$, mass of $3943 \pm 11 \pm 13$ MeV, 
can be any
of the following states $2^3P_0$, 3831 MeV, $2^3P_1$, 3938 MeV, or
$2^1P_1$, 3959 MeV; $Y(4260)$, mass of 4260 MeV, can be any of the following
states $3^3S_1$, 4164, $2^1D_2$, 4230 MeV, $2^3D_2$, 4230 MeV, $1^1G_4$, 
4253 MeV, or $1^3G_4$, 4260 MeV.
For charmonium $X(3872)$, mass of $3871.2 \pm 0.5$ MeV, the experimental data
favors the following quantum numbers 
$J^{PC} = 1^{++}, 2^{-+}$. For the potential considered here, the closest 
states with such quantum numbers are $2^3P_1$, 3938 MeV, and $2^3P_0$,
3832 MeV.

\begin{table}
\caption{Bottomonium spectra. The right columns report the difference, in MeV,
between theory and experimental numbers.}
\label{bottom}       
\begin{tabular}{rlr|rlr}
\hline\noalign{\smallskip}
 Theo  & Exp   &      & Theo  & Exp   &  \\
 (MeV) & (MeV) &      & (MeV) & (MeV) &  \\
\hline\noalign{\smallskip}
 \multicolumn{3}{l|}{$J^{PC} = 0^{-+}$} & 
                       \multicolumn{3}{l}{$J^{PC} = 1^{--} ~$ [$^3S_1$]}    \\
  9457 & $\eta_b (1S)$        &  157    &  9460 & $\Upsilon (1S)$ &     0   \\
 10018 & $\eta^\prime_b (2S)$ & ---     & 10023 & $\Upsilon (2S)$ &     0   \\
 10380 & ---                  & ---     & 10385 & $\Upsilon (3S)$ &    30   \\
 10721 & ---                  & ---     & 10727 & $\Upsilon (4S)$ &   148   \\
 11059 & ---                  & ---     & 11065 & $\Upsilon (11020)$ &  46  \\
\hline\noalign{\smallskip}
 \multicolumn{3}{l|}{[$1P$]} &  
 \multicolumn{3}{l}{[$2P$]} \\
  9983 & $\chi_{b2} (1P)$     &  71     & 10326 & $\chi_{b2} (2P)$ & 57  \\
  9941 & $\chi_{b1} (1P)$     &  48     & 10283 & $\chi_{b1} (2P)$ & 28  \\
  9894 & $\chi_{b0} (1P)$     &  35     & 10234 & $\chi_{b0} (2P)$ &  2   \\
  9955 & $h_b (1P)$           & ---     & 10296 & $h_b (2P)$       & ---  \\
\hline\noalign{\smallskip}
 \multicolumn{3}{l}{[$^3D_1$]} &  
 \multicolumn{3}{l}{} \\
  10159 & ---                & ---      & 10476 & $\Upsilon (4S)$ (?)& 103  \\
  10796 & $\Upsilon (10860)$ &  69      &       &                    &   \\
\hline\noalign{\smallskip}
 \multicolumn{3}{l}{[$^3D_2$]} &   10179 & $\Upsilon (1D)$ & 18  \\
\hline\noalign{\smallskip}
\end{tabular}
\end{table}

In table \ref{bottom} we report the low lying bottomonium states. 
The masses were shifted to reproduce the $\Upsilon (1S)$ experimental value. 

In conclusion, the non-relativistic potential $V_0$ is able to
explain the charmonium spectra with an error of less than 3\% ($\sim$ 100 MeV)
and the bottomonium spectra with an error of less than 1.5\% ($\sim$ 100 MeV).

%
\subsection{Leptonic Decays of Charmonium and Bottomonium}
\label{sec:3}

For the leptonic decays we follow the van Royen-Weisskopf \cite{RoWe67}
approach and assume that QCD corrections factorize in the calculation of
the widths, i.e.
\begin{eqnarray}
  \Gamma_{e^+ e^-} ( ^3S_1 ) & = &
        \frac{4 \, e^2_q \, \alpha^2}{M^2} 
        \, \left| R_{nS} (0) \right|^2 \, , \\
  \Gamma_{e^+ e^-} ( ^3D_1 ) & = &
    \frac{25 \, e^2_q \, \alpha^2}{2 \, m^4_q \, M^2} 
    \, \left| R^{(2)}_{nD} (0) \right|^2 \, ,
\end{eqnarray}
times QCD corrections; $e_q$ is the quark electric charge, $\alpha$ the fine
structure constant, $m_q$ the quark mass, $M$ the meson mass, $R_{nS} (0)$ the
S-wave meson radial wave function at the origin and $R^{(2)}_{nD} (0)$ the
D-wave meson second derivative of the radial wave function at the origin. To
avoid the problem of estimating the QCD corrections, in table \ref{leptonic}
we report the theoretical witdhs computed relative to $J / \psi (1S)$ for 
charmonium and $\Upsilon (1S)$ for bottomonium.

For charmonium, $\Gamma_{e^+e^-}$ for $\psi (2S)$ is slightly larger than the
experimental value. For bottomonium, the witdh for the $\Upsilon (2S)$ is
below the experimental number but the witdhs for $\Upsilon (3S)$ and 
$\Upsilon (4S)$ agree with the experimental figure within two standard 
deviations. For $\Upsilon (5S)$, the theoretical prediction is a factor of 3
higher than the experimental figure. For $D$-wave mesons, the theoretical 
predictions for the leptonic widths are much smaller than the measured 
$\Gamma_{e^+e^-}$. However, for $c \overline c$ if one considers mixing
betweem $S$ and $D$ mesons, it is possible to bring all widths quite close
to the experimental figures; see \cite{OlRi06} for details.

\begin{table}
\caption{Charmonium and Bottomonium Leptonic Widths in KeV. The experimental
figures are from \cite{PDG}. The limit for $Y(4260)$ is from \cite{Yee}.}
\label{leptonic}       
\begin{tabular}{l|lrl|lrl}
\hline\noalign{\smallskip}
          &             &  Theo  & Exp    & 
                        &  Theo  & Exp    \\
\hline\noalign{\smallskip}
 $2^3S_1$  &  $\psi (2S)$  \hspace{-0.35cm}   & 2.84  & $2.48(6)$ &
              $\Upsilon (2S)$ \hspace{-0.35cm} & 0.426 & $0.612(11)$ \\
 $3^3S_1$  &  $Y(4260)$    \hspace{-0.35cm}   & 2.16    & $<0.40$ &
              $\Upsilon (3S)$ \hspace{-0.35cm} & 0.356 & $0.443(8)$ \\
 $4^3S_1$  &  ---             & ---    & --- &
              $\Upsilon (4S)$ \hspace{-0.35cm} & 0.335 & $0.272(29)$ \\
 $5^3S_1$  &  ---             & ---    & --- &
              $\Upsilon (5S)$ \hspace{-0.35cm} & 0.311 & $0.130(30)$ \\
\hline\noalign{\smallskip}
\end{tabular}
\end{table}

%
\subsection{E1 Electromagnetic Transitions}
\label{sec:4}

For the computation of the E1 electromagnetic charmonium transitions we
follow \cite{KwRo88}. Table \ref{E1velhos} report the theoretical estimates
of the decay widths known experimentaly. On overall, the agreement between
theory and experiment is good. The exception being the transitions involving
the scalar meson $\chi_{c0} (1P)$. Remember that for the meson spectra, the
larger deviations from the experimental numbers occured for the scalar mesons.

\begin{table}
\caption{E1 electromagnetic charmonium widths in KeV. In the calculation, for
the meson mass it was used the experimental measured mass. Experimental
numbers are from \cite{PDG}.}
\label{E1velhos}       
\begin{tabular}{llrr}
\hline\noalign{\smallskip}
 & & Theo & Exp \\
\hline\noalign{\smallskip}
 $\psi (2S) ~ \longrightarrow$ & $\chi_{c2} (1P) + \gamma$ & 
                      $30$ & $27 \pm 2$ \\
                               & $\chi_{c1} (1P) + \gamma$ & 
                      $43$ & $29 \pm 2$ \\
                               & $\chi_{c0} (1P) + \gamma$ & 
                      $51$ & $31\pm 2$ \\
\hline\noalign{\smallskip}
 $\chi_{c2} (1P) ~ \longrightarrow$ & $J / \psi (1S) + \gamma$ & 
                      $414$ & $416 \pm 32$ \\
 $\chi_{c1} (1P) ~ \longrightarrow$ & $J / \psi (1S) + \gamma$ & 
                      $308$ & $317 \pm 25$ \\
 $\chi_{c0} (1P) ~ \longrightarrow$ & $J / \psi (1S) + \gamma$ & 
                      $146$ & $135 \pm 15$ \\
\hline\noalign{\smallskip}
 $\psi (3770) (1^3D_1) ~ \longrightarrow$ & $\chi_{c2} (1P) + \gamma$ & 
                      $4$ & $< 21$ \\
                                          & $\chi_{c1} (1P) + \gamma$ & 
                      $110$ & $75 \pm 18$ \\
                                          & $\chi_{c0} (1P) + \gamma$ & 
                      $359$ & $172 \pm 30$ \\
\hline\noalign{\smallskip}
\end{tabular}
\end{table}

Given the good agreement for the known experimental radiative transitions
of the charmonium, for the theoretical states which can describe the new
charmonium states we have computed their E1 electromagnetic widths. In table
\ref{E1novos} we report those widths which are larger than $\sim 40$ KeV. In
principle, the electromagnetic transitions can allow to distinguish not only
the various theoretical models but also the difference meson states.

\begin{table}
\caption{E1 electromagnetic charmonium widths, in KeV, for the new charmonium
states. In the calculation, for the meson mass it was used the experimental 
measured mass. Only transitions to wellknown states are considered.}
\label{E1novos}       
\begin{tabular}{llrr}
\hline\noalign{\smallskip}
 & & $E_\gamma$ & $\Gamma$ \\
\hline\noalign{\smallskip}
 $X(3872) [2^3P_1] ~ \longrightarrow$ 
          & $J / \psi (1S) + \gamma$ & $697$ & $48$ \\
          & $\psi (2S) + \gamma$     & $181$ & $63$ \\
\hline\noalign{\smallskip}
\multicolumn{4}{l}{$X(3940)$, $Y(3940)$} \\
 $2^3P_1$ or $2^3P_0   ~ \longrightarrow$ 
          & $J / \psi (1S) + \gamma$ & $755$ & $61$  \\
          & $\psi (2S) + \gamma$     & $249$ & $165$ \\
 $2^3P_0 ~ \longrightarrow$ 
          & $\psi (3770) + \gamma$   & $168$ & $70$  \\
 $2^1P_1 ~ \longrightarrow$ 
          & $\eta_c (1S) + \gamma$          & $845$ & $86$  \\
          & $\eta^\prime_c (2S) + \gamma$   & $293$ & $271$  \\
\hline\noalign{\smallskip}
 $Z(3930) [2^3P_2] ~ \longrightarrow$ 
          & $J / \psi (1S) + \gamma$ & $744$ & $59$  \\
          & $\psi (2S) + \gamma$     & $235$ & $140$ \\
\hline\noalign{\smallskip}
\end{tabular}
\end{table}

To finish, in table \ref{E1bb} we report the bottomonium E1 
electromagnetic transitions.

\begin{table}
\caption{E1 electromagnetic bottomonium widths in KeV. The experimental numbers
are from \cite{bbE1}.}
\label{E1bb}       
\begin{tabular}{lllr}
\hline\noalign{\smallskip}
 & &  Theo & Exp \\
\hline\noalign{\smallskip}
 $\Upsilon (2S) ~ \longrightarrow$ 
          & $\chi_{c2} (1P) + \gamma$ & $3.30 \pm 0.04$ & $2.21 \pm 0.16$ \\
          & $\chi_{c1} (1P) + \gamma$ & $3.21 \pm 0.04$ & $2.11 \pm 0.16$ \\
          & $\chi_{c0} (1P) + \gamma$ & $2.10 \pm 0.02$ & $1.14 \pm 0.16$ \\
\hline\noalign{\smallskip}
 $\Upsilon (3S) ~ \longrightarrow$ 
          & $\chi_{c2} (2P) + \gamma$ & $3.22$ & $2.95 \pm 0.21$ \\
          & $\chi_{c1} (2P) + \gamma$ & $2.95$ & $2.71 \pm 0.20$ \\
          & $\chi_{c0} (2P) + \gamma$ & $1.82$ & $1.26 \pm 0.14$ \\
\hline\noalign{\smallskip}
\end{tabular}
\end{table}
%
%

\section*{Acknowledgments}

R. A. M. acknowledges F.C.T. for financial support, grant SFRH/BD/8736/2002.
This work was partly supported by FCT under contract POCI/FP/63436/2005.

%
%

\end{document}